\documentstyle[preprint,aps,epsf,amstex]{revtex}
\begin{document}
\draft
%
%
%
%
\title{
Dephasing in
Disordered Conductors due to Fluctuating Electric Fields}
\author{Axel V\"{o}lker and Peter Kopietz}
\address{
Institut f\"{u}r Theoretische Physik der Universit\"{a}t G\"{o}ttingen,
Bunsenstrasse 9, D-37073 G\"{o}ttingen, Germany}
\date{June 18, 1999}
\maketitle
\begin{abstract}

We develop a novel
eikonal expansion for
the Cooperon to study
the effect of  space- and time-dependent
electric fields
on the dephasing rate of disordered conductors.
For randomly fluctuating fields with arbitrary covariance
we derive a general expression for the 
dephasing rate
which is free of infrared divergencies in reduced dimensions. 
For time-dependent  external fields
with finite wavelength and sufficiently small
amplitude
we show that the dephasing rate is
proportional to the square root of
the electromagnetic power coupled into the system,
in agreement with data by
Wang and Lindelof [Phys. Rev. Lett. {\bf{59}}, 1156 (1987)].

\end{abstract}
\narrowtext

\section{Introduction}

The dephasing time $\tau_{\phi}$
of a particle in a quantum system is the time
over which its wave-function  maintains
phase coherence. For diffusive systems 
in dimensions $d=1$ and $d=2$ the dephasing time
can be obtained experimentally
from the weak-localization correction
to the conductivity in a magnetic field.
Recently Mohanty, Jariwala and Webb\cite{Mohanty97}
used this method a measure $\tau_{\phi}$ as a function
of the temperature $T$ in 
gold wires. They found that at sufficiently
low temperatures $\tau_{\phi}$ approaches a finite value.
This is in disagreement with the generally accepted
point of view\cite{Imry97} that in thermal equilibrium $\tau_{\phi}$
should diverge as $T^{-p}$, $p > 0$, for $T \rightarrow 0$.
A proposal that the observed saturation of
$\tau_{\phi}$ is  an intrinsic ground-state property
of interacting electrons in a random potential\cite{Mohanty97b}
has been heavily criticized\cite{Aleiner98}. 
Until now, there is no general agreement 
on the microscopic mechanism for the observed saturation
of $\tau_{\phi}$ at low temperatures.

One possible reason for this saturation
might be external microwave radiation that is
unintentionally 
coupled into the  system\cite{Altshuler98}. 
According to Ref.\cite{Altshuler81}
the dephasing rate $1/ \tau_{\rm AC} $
due to a time-dependent but spatially constant
electric field is proportional to the microwave power
$P$ absorbed by the system for small $P$, 
and crosses over to a $P^{1/5}$-law for larger $P$.
Experimentally the effect of microwaves on weak localization
has been studied by several authors
\cite{Wang87,Vitkalov88,Liu91,Webb99}. 
Most of the data by Wang and Lindelof\cite{Wang87}
can be fitted with $1/ \tau_{\rm AC} \propto P^{1/2}$.
Below we shall offer a simple explanation
for this behavior.  More recent data 
by Webb {\it{et al.}}\cite{Webb99}  are consistent with
a $P^{1/5}$-law at high powers $P$, but
in this experiment the microwaves seem to heat the sample,
an effect which has not been taken into account 
in Ref.\cite{Altshuler81}, and which we will neglect as well.

Following Altshuler {\it{et al.}}\cite{Altshuler81,Altshuler82}, we define
the dephasing rate in the diffusive regime via
the weak localization correction $\delta \sigma$
to the static conductivity.
In the presence of an external electric field
$\delta \sigma$ can be written 
as\cite{Altshuler81,Altshuler82,Voelker99}
 \begin{equation}
 \delta \sigma
  = - \frac{ \sigma_0 }{\pi \nu_d } 
 \int_{\tau_{\rm el} }^{\infty} d t 
 \lim_{T_0 \rightarrow \infty}
 \frac{1}{2 T_0}
 \int_{ - T_0 }^{T_0} dt_0
 {\cal{C}} ( {\bf{r}} , {\bf{r}} , 
 \frac{t}{2} , - \frac{t}{2} , t_0 )
 \; ,
 \label{eq:sigmane}
 \end{equation}
where $\sigma_0$ is the Drude conductivity,
$\nu_d$ is the $d$-dimensional density of states,
and
$\tau_{\rm el}$ is the momentum relaxation time.
The Cooperon ${\cal{C}} $ 
satisfies\cite{Voelker99,Eiler84,Vavilov99}
 \begin{eqnarray}
 \left[  \partial_t  + D {\hat{\bf{P}}}_{\bf{r}}^2  + \Gamma_0
 + i [
 V ( {\bf{r}} , t_0 + t  )  -
 V ( {\bf{r}} , t_0 - t  )  ] \right]
 \nonumber
 \\
 & & \hspace{-55mm} \times
 {\cal{C}} ( {\bf{r}} , {\bf{r}}^{\prime} , t , t^{\prime} , t_0 )
 = \delta ( {\bf{r}} - {\bf{r}}^{\prime} )
 \delta ( t - t^{\prime} )
 \; , 
 \label{eq:Cooperondef}
 \end{eqnarray}
where $\hat{\bf{P}}_{\bf{r}} = - i \nabla_{\bf{r}}$ is the
momentum operator,
$D$ is the diffusion coefficient, and
the phenomenological cutoff 
$\Gamma_0$ describes dephasing due to processes that are not
explicitly treated in our calculation, such as 
inelastic electron-phonon scattering
or electron-electron scattering with
energy-transfers
$ | \omega |
{ \raisebox{-0.5ex}{$\; \stackrel{>}{\sim} \;$}}  T$.
The potential $V ( {\bf{r}} , t )$ is
related to the longitudinal electric field via
$ e {\bf{E}} ( {\bf{r}} , t ) =  \nabla_{\bf{r}}  V ( {\bf{r}} , t )$,
where $-e$ is the charge of the electron.
For $\Gamma_0 = V = 0$ 
the $t$-integration in Eq.(\ref{eq:sigmane})
diverges in $ d \leq 2$ at the upper limit, but for finite $\Gamma_0$ or $V$
the integration is cut off at some finite time,
the so-called dephasing time.

\section{Eikonal expansion}

Formally Eq.(\ref{eq:Cooperondef}) looks like the
differential equation for the imaginary-time single-particle Green's function
of an electron in a fluctuating external potential
\begin{equation}
 {V}_{t_0} ( {\bf{r}} , t )   = 
 i [ V ( {\bf{r}} , t_0 + t) -
  V ( {\bf{r}} , t_0 - t) ].
\end{equation}
The real time version of this problem
has been discussed extensively in the
quantum field theory literature\cite{Svidzinskii57}.  
By means of a simple modification of the
method developed by E. S. Fradkin\cite{Svidzinskii57} (see also 
Ref.\cite{Kopietz96}) we obtain
the solution of Eq.(\ref{eq:Cooperondef}) 
in the following form\cite{Voelker99}
 \begin{eqnarray}
 {\cal{C}} ( {\bf{r}} , {\bf{r}}^{\prime} , t , t^{\prime} , t_0 ) &  = &
 \Theta ( t - t^{\prime} )
 \int \frac{ d {\bf{k}} }{ ( 2 \pi )^d}
 e^{i {\bf{k}} \cdot ( {\bf{r}} - {\bf{r}}^{\prime} ) }
 \nonumber
 \\
 & \times &
  e^{ 
  -    (\Gamma_0  + D {\bf{k}}^2) ( t - t^{\prime} )
  - F ( {\bf{k}} , t - t^{\prime} ;
 {\bf{r}} , t ) }
 \; ,
 \label{eq:Gsol}
 \end{eqnarray}
where the function $F$ satisfies the eikonal equation
 \begin{eqnarray}
 \left[  \partial_{\tau }
 + \partial_t  + D ( \hat{\bf{P}}_{\bf{r}}^2 
 + 2 {\bf{k}} \cdot \hat{\bf{P}}_{\bf{r}} ) \right]
 F ( {\bf{k}} , \tau ; {\bf{r}} , t ) & =  &
 \nonumber
 \\
 & & \hspace{-60mm}
 {V}_{t_0} ( {\bf{r}} , t )   + D
 \left[ \hat{\bf{P}}_{\bf{r}}
 F ( {\bf{k}} , \tau ; {\bf{r}} , t )  \right]^2 
 \label{eq:F}
 \; ,
 \end{eqnarray}
with boundary condition 
 $ F ( {\bf{k}} , 0 ; {\bf{r}}, t ) = 0$.
Although in general Eq.(\ref{eq:F}) cannot be solved exactly,
we can easily obtain an expansion 
of $F$ in powers of the potential
${V}_{t_0} ( {\bf{r}} , t )$. Setting
\begin{equation} 
F ( {\bf{k}} , \tau ; {\bf{r}}, t ) 
= \sum_{n=1}^{\infty} F_n ( {\bf{k}} , \tau ; {\bf{r}}, t ) 
\; , 
\end{equation}
where $F_n$ involves
by definition $n$ powers of ${V}_{t_0}$,
successive terms may be 
calculated recursively\cite{Voelker99,Svidzinskii57,Kopietz96}.
For our purpose we need only the first two
terms.
The linear term is 
 \begin{eqnarray}
 F_1 ( {\bf{k}} , \tau ; {\bf{r}} , t )
   & = &  \int \frac{d {\bf{q}} d \omega }{(2 \pi )^{d+1}}
  e^{ i ( {\bf{q}} \cdot {\bf{r}} - \omega t )}
  {V}_{t_0} ( {\bf{q}} , \omega ) 
  \nonumber
  \\
  & & \times
   \frac{ 1 - e^{ - [D ( {\bf{q}}^2 + 2 {\bf{k}} \cdot {\bf{q}} ) 
  - i \omega ]
  \tau } }{D ( {\bf{q}}^2 + 2 {\bf{k}} \cdot {\bf{q}} ) -
  i \omega}
  \label{eq:F1sol}
  \; .
  \end{eqnarray}
Here
\begin{equation}
\label{Vt0}
{V}_{t_0} ( {\bf{q}} , \omega )
=
 i [
 e^{- i \omega t_0 } V ( {\bf{q}} , \omega ) - e^{i \omega t_0 }
 V ( {\bf{q}} , - \omega ) ]
\end{equation}  
is the Fourier transform  of
${V}_{t_0} ( {\bf{r}} , t )$, 
where
\begin{equation}  
V ( {\bf{q}} , \omega ) = \int d {\bf{r}} \int dt
 e^{-i ( {\bf{q}} \cdot {\bf{r}} - \omega t ) }
 V ( {\bf{r}} , t ) \; .
\end{equation}
From Eq.(\ref{eq:F}) we find that
the quadratic term is related to the linear one via
\begin{equation}
\label{F1F2}
F_2 ( {\bf{k}} , \tau ; {\bf{r}} , t ) = - \int_0^{\tau} d\tau'
e^{(D{\bf \hat P}_{\bf r}^2 +2D{\bf k\cdot \hat P}_{\bf r}
  +\partial_t)(\tau'-\tau)} D\left[{\bf \hat P}_{\bf r}F_1({\bf
    k},\tau' ; {\bf r} , t) \right]^2 \; . 
\end{equation}
The explicit calculation yields
 \begin{eqnarray}
 F_2 ( {\bf{k}} , \tau ; {\bf{r}} , t )
   & = &  
   \int \frac{d {\bf{q}}_1 d \omega_1}{(2 \pi )^{d+1}}
   \int \frac{d {\bf{q}}_2 d \omega_2 }{(2 \pi )^{d+1}}
  \nonumber
  \\
  & & \hspace{-23mm} \times
  e^{ i[ ( {\bf{q}}_1 + {\bf{q}}_2 ) \cdot {\bf{r}} - 
  ( \omega_1 + \omega_2 ) ]t }
  {V}_{t_0} ( {\bf{q}}_1 , \omega_1 ) 
  {V}_{t_0} ( {\bf{q}}_2 , \omega_2 ) 
  \nonumber
  \\
  & & \hspace{-23mm} \times
  \frac{ 
  ( D {\bf{q}}_1 \cdot {\bf{q}}_2  )
  e^{ - [ D ( ( {\bf{q}}_1 + {\bf{q}}_2 )^2 
  + 2 {\bf{k}} \cdot ( {\bf{q}}_1 + {\bf{q}}_2 ) ) - i
  ( \omega_1 + \omega_2 ) ] \tau }
  }
  { 
  \left[
  D ( {\bf{q}}_1^2 + 2 {\bf{k}} \cdot {\bf{q}}_1 ) - i \omega_1 
  \right]
  \left[
  D ( {\bf{q}}_2^2 + 2 {\bf{k}} \cdot {\bf{q}}_2 ) - i \omega_2 
  \right]}
  \nonumber
  \\
  &  &  \hspace{-23mm} \times
  \left\{
  \frac{
  e^{  [ 
  D ( ( {\bf{q}}_1 + {\bf{q}}_2 )^2 
  + 2 {\bf{k}} \cdot ( {\bf{q}}_1 + {\bf{q}}_2 ) ) - i
  ( \omega_1 + \omega_2 ) ] \tau } - 1}
  {
  D ( ( {\bf{q}}_1 + {\bf{q}}_2 )^2 
  + 2 {\bf{k}} \cdot ( {\bf{q}}_1 + {\bf{q}}_2 ) ) - i
  ( \omega_1 + \omega_2 ) }
  \nonumber
  \right.
  \\
  &  &  \hspace{-18mm}
  + \frac{ e^{ 2 D {\bf{q}}_1 \cdot {\bf{q}}_2  \tau } -1}
  {
  2 D {\bf{q}}_1 \cdot {\bf{q}}_2  }
  - \frac{
  e^{  [ 
  D (  {\bf{q}}_1^2 
  + 2 {\bf{k}} \cdot  {\bf{q}}_1  + 2 {\bf{q}}_1 \cdot {\bf{q}}_2 ) - i
   \omega_1 ] \tau } - 1}
  {
  D (  {\bf{q}}_1^2 
  + 2 {\bf{k}} \cdot  {\bf{q}}_1  + 2 {\bf{q}}_1 \cdot {\bf{q}}_2  ) - i
   \omega_1  }
   \nonumber
   \\
   & & \hspace{-18mm} 
   \left.
  -
  \frac{
  e^{  [ 
  D (  {\bf{q}}_2^2 
  + 2 {\bf{k}} \cdot  {\bf{q}}_2  + 2 {\bf{q}}_2 \cdot {\bf{q}}_1 ) - i
   \omega_2 ] \tau } - 1}
  {
  D (  {\bf{q}}_2^2 
  + 2 {\bf{k}} \cdot  {\bf{q}}_2  + 2 {\bf{q}}_2 \cdot {\bf{q}}_1  ) - i
   \omega_2  }
  \right\}
  \; .
  \label{eq:F2sol}
  \end{eqnarray}
Because we have made the diffusion approximation,
the momentum-integrations in
Eqs.(\ref{eq:F1sol},\ref{eq:F2sol}) 
are restricted to
$|{\bf q}|, |{\bf q}_1|, |{\bf q}_2|<  1/ \ell  \equiv  1 / ( v_F \tau_{\rm el})$, 
and the
frequency-integrals to $|\omega|, |\omega_1|, |\omega_2| < 1/\tau_{ \rm el}$. 
Here $\ell $ is the
elastic mean free path  and $v_F$ is the Fermi velocity.
For
brevity we have not explicitly written out these cutoffs  
in the above expressions.

\section{Dephasing due to random fields}

\subsection{General case}

We now assume that the potential $V ( {\bf{r}} , t )$
is a random function with zero average and general covariance
 \begin{equation}
 \langle V ( {\bf{q}} , \omega ) V ( {\bf{q}}^{\prime} ,
 \omega^{\prime} ) \rangle =
  (2 \pi )^{d+1}  \delta ( {\bf{q}} + {\bf{q}}^{\prime} )
 \delta ( \omega + \omega^{\prime} )  
 g ( {\bf{q}} , \omega )
 \; ,
 \label{eq:corphi}
 \end{equation}
where $\langle \ldots \rangle$ denotes averaging
over the probability distribution of
$V$. The dephasing rate is then defined
in terms of the average
$\langle \delta \sigma \rangle$ 
of Eq.(\ref{eq:sigmane}).
Using the fact that after  the averaging the Cooperon 
is independent of the
time $t_0$, we obtain from
Eqs.(\ref{eq:sigmane},\ref{eq:Gsol})
 \begin{equation}
 \langle \delta \sigma \rangle
 = - \frac{\sigma_0 }{\pi \nu_d} \int_{\tau_{\rm el}}^{\infty} dt 
 \int \frac{ d {\bf{k}}}{
 ( 2 \pi )^d} e^{ - ( \Gamma_0 + D {\bf{k}}^2 ) t
 - \Gamma ( {\bf{k}} , t ) }
 \; ,
 \label{eq:sigmawl2}
 \end{equation}
where
 \begin{equation}
 \Gamma ( {\bf{k}} , t )  =
  - \ln \langle e^{ - F ( {\bf{k}},  t ; {\bf{r}} , t/2 ) } \rangle
  \; .
  \end{equation}
We now perform a linked cluster expansion of
$\Gamma ( {\bf{k}} , t ) $ in powers of the
correlator $g ( {\bf{q}} , \omega)$.
To first  order  we find 
 \begin{equation}
\Gamma ( {\bf{k}} , t )  \approx   
\Gamma_1 ( {\bf{k}} , t ) 
+
\Gamma_2 ( {\bf{k}} , t ) 
\; ,
\end{equation}
where
 \begin{eqnarray}
 \Gamma_1 ( {\bf{k}} , t ) 
  & =  &
 - \frac{1}{2} \langle F_1^2 ( {\bf{k}} , 
 t ; {\bf{r}} , t/2 ) \rangle
 \; ,
 \\
 \Gamma_2 ( {\bf{k}} , t ) 
  & = &
 \langle F_2 ( {\bf{k}} , t ; {\bf{r}} , t/2 ) \rangle
 \; .
 \end{eqnarray}
Introducing the notation
 $E_{\bf{k}} ( {\bf{q}} ) = D ( {\bf{q}}^2 + 2 {\bf{k}} \cdot {\bf{q}} )$
we obtain
 \begin{eqnarray}
 \Gamma_1 ( {\bf{k}} , t )  
   =  
 \int \frac{ d {\bf{q}}  d \omega }{(2 \pi )^{d+1}}
 g ( {\bf{q}},  \omega )
 & &
 \nonumber
 \\
 &  & \hspace{-41mm}  \times \left\{
  \frac{ 1 - e^{ - ( E_{\bf{k}} ( {\bf{q}} ) - i \omega  ) 
  t }}{E_{\bf{k}} ( {\bf{q}} ) - i \omega  }
  \; \; 
  \frac{ 1 - e^{ - ( E_{\bf{k}} ( - {\bf{q}} ) + i \omega  ) 
  t }}{E_{\bf{k}} ( - {\bf{q}} ) + i \omega  }
  \right.
  \nonumber
  \\
  & &
  \hspace{-41mm}  \left.
  - e^{- i \omega t }
  \frac{ 1 - e^{  - ( E_{\bf{k}} ( {\bf{q}} ) - i \omega )  t}  
  }{E_{{\bf{k}}} ( {\bf{q}} ) - i \omega }
  \; \; 
  \frac{ 1 - e^{  - ( E_{\bf{k}} ( -{\bf{q}} ) - i \omega )  t }
  }{E_{\bf{k}} ( - {\bf{q}} ) - i \omega }
   \right\}
  \; ,
  \label{eq:Qares}
  \end{eqnarray}
 \begin{eqnarray}
 \Gamma_2 ( {\bf{k}} , t )  & = &  
  \int \frac{ d {\bf{q}} d \omega }{(2 \pi )^{d+1}}
 g ( {\bf{q}},  \omega )
 \frac{ 2 D {\bf{q}}^2}{
  E_{\bf{k}} ( {\bf{q}} ) - i \omega } 
 \nonumber
 \\
 &  & \hspace{-14mm}  \times \left\{
 \frac{1}{
  E_{\bf{k}} ( - {\bf{q}} )  + i \omega  }
 \left[
 t 
 + \frac{ 1 - e^{- 2 D {\bf{q}}^2 t }}{2 D {\bf{q}}^2 }
 - \frac{ 1 - e^{ - ( E_{\bf{k}} ( {\bf{q}} ) - i \omega ) t }}{
 E_{\bf{k}} ( {\bf{q}} ) - i \omega }
 - \frac{ 1 - e^{ - ( E_{\bf{k}} ( - {\bf{q}} ) + i \omega ) t }}{
 E_{\bf{k}} ( -{\bf{q}} ) + i \omega }
 \right]
 \right.
 \nonumber
 \\
 &  & \hspace{-12mm}  \left.
 - \frac{1}{ 
  E_{\bf{k}} ( - {\bf{q}} )  - i \omega  }
 \left[
 \frac{\sin ( \omega t )}{ \omega }
 + e^{i \omega t}
 \left[
 \frac{ 1 - e^{- 2 D {\bf{q}}^2 t }}{2 D {\bf{q}}^2 }
 - \frac{ 1 - e^{ - ( E_{\bf{k}} ( {\bf{q}} ) + i \omega ) t }}{
 E_{\bf{k}} ( {\bf{q}} ) + i \omega }
 - \frac{ 1 - e^{ - ( E_{\bf{k}} ( - {\bf{q}} ) + i \omega ) t }}{
 E_{\bf{k}} ( - {\bf{q}} ) + i \omega }
 \right]
 \right]
 \right\}
  \; .
  \label{eq:Qbres}
  \end{eqnarray}

\subsection{Nyquist noise}

As a special case 
let us assume that the potential $V$ 
is generated by
equilibrium fluctuations  of the
electric field due to the thermal motion of the
electrons (Nyquist noise).
Then $g ( {\bf{q}} , \omega )$ is determined by
the fluctuation-dissipation theorem,
which implies\cite{Altshuler82}
\begin{equation}
 g ( {\bf{q}} , \omega ) = - f_{\bf{q}} \coth ( \frac{\omega}{2T} )
 {\rm Im} \epsilon^{-1} ( {\bf{q}} , \omega ) \; ,
\end{equation}
where $f_{\bf{q}}$ is the Fourier transform of the bare
Coulomb interaction, and the
dielectric function of the system is
in the diffusive regime given by
 \begin{equation}
 \epsilon ( {\bf{q}} , \omega ) = 1
 + f_{\bf{q}} \nu_d \frac{ D {\bf{q}}^2  }{ D {\bf{q}}^2 - i \omega }
 \; .
 \label{eq:epsdef}
 \end{equation}
The corresponding dephasing rate in
$d=1,2,3$ was calculated by Altshuler {\it{et al.}}\cite{Altshuler82}
with the help of a Feynman path-integral  representation
of the solution of  Eq.(\ref{eq:Cooperondef}).
It is instructive to reproduce
the results of Ref.\cite{Altshuler82} within
our eikonal expansion, treating $d$ as a continuous parameter.
It is important to note that
in the derivation of Eq.(\ref{eq:Cooperondef}) the
electromagnetic field is treated classically\cite{Altshuler82,Vavilov99}, 
so that our calculation
takes only into account
low-frequency Fourier components of the screened
Coulomb interaction, with
$| \omega | 
{ \raisebox{-0.5ex}{$\; \stackrel{<}{\sim} \;$}}  T$.
The correlator $g ( {\bf{q}} , \omega )$ in Eqs.(\ref{eq:Qares},\ref{eq:Qbres})
can then be approximated by
 \begin{equation}
g ( {\bf{q}} , \omega ) \approx \Theta ( T - | \omega | )
\frac{ 2 T}{ \nu_d D {\bf{q}}^2 }
\; ,
\end{equation}
Because the momentum integrals in the following analysis
will be infrared {\it{and}} ultraviolet convergent, 
no further cutoffs are needed
in our approach.

The dephasing rate is determined
by the long-time behavior of
$\Gamma ( t ) \equiv \Gamma ( {\bf{k}} = 0 , t )$.
In $d < 2$ we find 
for $t  \gg 1/T$
\begin{equation}
\Gamma ( t ) \sim  C_d
\frac{ T t^{2 - \frac{d}{2}}  }{\nu_d D^{\frac{d}{2}}}  
\;  ,
 \end{equation}
where the numerical constant $C_d$ is given by
 \begin{equation}
C_d =  \frac{2^{3- {d}}} 
{\pi^{\frac{d}{2}} (2-d) (4-d) } \; .
 \end{equation}
Because $\Gamma ( t ) $ grows for large $t$ 
faster than linear, the term $\Gamma_0 t$ in Eq.(\ref{eq:sigmawl2})
is negligible, and the dephasing rate may be defined by
 $\Gamma ( \tau_{\phi} ) = 1$, which yields
\begin{equation}
\frac{1}{ \tau_{\phi}} = 
\left[\frac { C_d T }{  \nu_d D^{d/2}  } \right]^{\frac{2}{4-d}} \; .
\end{equation}
In $d =2$ we find 
for $t \gg 1/T$
 \begin{equation}
 \Gamma ( t ) \sim \frac{ \ln ( T t )  Tt}{ 2 \pi \nu_2 D } 
 \; ,
\end{equation}
so that $\tau_{\phi}$ satisfies
\begin{equation}
 \frac{1}{\tau_{\phi} } = \frac{ \ln ( T \tau_{\phi} ) T} 
 {2 \pi \nu_2 D}  \; .
\end{equation}
Keeping in mind that in a good metal
$2 \pi \nu_2 D  \gg 1$,
this implies to leading order
\begin{equation}
\frac{1}{\tau_{\phi} } =  \frac{\ln ( 2 \pi \nu_2 D )  {T}}{2 \pi
  \nu_2 D} \; ,
\end{equation}
in agreement with Ref.\cite{Altshuler82}.
Finally, in $d > 2$
the long-time behavior of  
$\Gamma ( t )$ is dominated by
the term proportional to $t$ in the second line of Eq.(\ref{eq:Qbres}). 
The total dephasing rate
in $d > 2$ can then be written as
 \begin{eqnarray}
 \frac{1}{\tau_\phi} & = & \Gamma_0 +
 \int \frac{ d {\bf{q}} d \omega }{(2 \pi )^d}
 g ( {\bf{q}},  \omega ) 
 \left[
 \frac{1}{\pi} \frac{ D {\bf{q}}^2 }{ ( D {\bf{q}}^2 )^2
 + \omega^2 }
 \right]
 \nonumber
 \\
 & = & 
 \Gamma_0 + 
 \tilde{C}_d
 {T^{\frac{d}{2}} } / ( {\nu_d D^{\frac{d}{2}}})
 \; ,
 \label{eq:Qb2}
 \end{eqnarray}
 where
\begin{equation}
\tilde{C}_d = \frac{2^{2-d}}{ \pi^{\frac{d}{2}} 
 (d-2)  \Gamma ( \frac{d}{2} ) \sin ( \frac{d}{4} \pi ) } \; .
\end{equation} 
Note that for $d \rightarrow 2$ both prefactors ${C}_d$ and 
$\tilde{C}_d$ diverge as 
$\pi^{-1} | d-2 |^{-1}$, signalling
logarithmic corrections in $d=2$.
The term in the braces of Eq.(\ref{eq:Qb2})
is the dynamic structure factor
of the diffusing electrons 
in the  regime 
$ | \omega | { \raisebox{-0.5ex}{$\; \stackrel{<}{\sim} \;$}}  T$,
where the detailed balance factor $e^{- \omega / T}$ 
can be replaced by unity.
This term agrees with the semiclassical 
dephasing rate derived by Chakravarty and Schmid\cite{Chakravarty86},
see also Ref\cite{Imry99}.
It should be kept in mind,
however, that Eq.(\ref{eq:Qb2})   
is only valid in $d > 2$, where
the integral is finite.
Diagrammatically Eq.(\ref{eq:Qb2}) ignores
vertex corrections in the Bethe-Salpeter equation
for the Cooperon, which become important
in $d \leq 2$. In this case Eq.(\ref{eq:Qb2})
should be replaced by  the more general expressions
(\ref{eq:Qares},\ref{eq:Qbres}), which
are free of infrared divergencies.


Our $\Gamma ( t )$ corresponds precisely
to the function $f_d ( t )$ introduced 
in a recent preprint by Golubev and Zaikin\cite{Golubev99}.
To make contact with this work, 
let us {\it{assume}} that
Eq.(\ref{eq:Cooperondef}) remains valid 
for potentials $ V ( {\bf{r}}, t )$ due to Nyquist noise with
frequencies in the range
$| \omega |  < 1/\tau_{\rm el} $, so that
in Eqs.(\ref{eq:Qares},\ref{eq:Qbres}) 
we may approximate
$g ( {\bf{q}} , \omega ) \approx  \Theta
(\tau_{\rm el}^{-1}  -  |\omega | ) \omega \coth ( \frac{\omega}{2T} ) 
/ ( \nu_d D {\bf{q}}^2 )$.  Then we obtain
in the quantum regime $T t \ll 1$
in quasi $1d$ to leading order
 $\Gamma ( t ) \sim 
  \sqrt{2} \pi^{-1}   \tau^{-{1}/{2}}_{\rm el} t/ (\nu_1 D^{\frac{1}{2}})$, 
which agrees 
with the leading term in the expansion of
the function $f_1 ( t )$ given in Eq.(28) of Ref.\cite{Golubev99}.
Moreover, also the subleading corrections to $f_1 ( t )$ given 
by Golubev and Zaikin\cite{Golubev99}
can be obtained within our eikonal expansion\cite{Voelker99}.
Hence, at least in $d=1$ we can completely reproduce the behavior of the
function $f_d ( t )$ 
discussed in Ref.\cite{Golubev99} 
{\it{if we assume that the 
differential equation (\ref{eq:Cooperondef}) for the
Cooperon remains valid when the classical potential is
replaced by a quantum field mediating the
Coulomb interaction.}}  
This replacement 
has been claimed by Golubev and Zaikin to be consistent with a fully
quantum mechanical calculation. It  
is the origin for the discrepancies between
their work and Ref.\cite{Altshuler82}.

\section{
Dephasing due to external fields}

In  the experiments\cite{Wang87,Webb99} microwaves
are coupled into the system via an  antenna
attached to a suitable waveguide, such that alternating
longitudinal currents are induced in the sample.
Keeping in mind that the precise way in which the microwaves couple
into the system is not known, we assume for simplicity that
the electrons feel a longitudinal electric field
of the form
\begin{equation}
 E ( {\bf{r}} , t ) =
 {E}_0 
 \cos ( {\bf{q}}_0 \cdot {\bf{r}}
 - \omega_0 t ) \; ,
\label{Ewave}
\end{equation}
where
$\omega_0$ is the microwave frequency and
the wave-vector ${\bf{q}}_0$ depends on the
geometry of the waveguide and the antenna.
The corresponding
potential
in Eq.(\ref{eq:Cooperondef}) is
\begin{equation}
V ( {\bf{r}} , t ) = V_0 \sin ( {\bf{q}}_0 \cdot {\bf{r}} 
- \omega_0 t ) \; ,
\end{equation}
with
$E_0 = q_0 V_0 / e$.
Let us emphasize that this is the total screened potential, which
is the sum of the external potential and the
induced potential.
Of course, the Maxwell equation inside a metal contains
a dissipative term, so that the field inside the metal 
is not given by a simple propagating wave\cite{Ziman72}.
In  general we expect that the
field distribution inside the metal  
depends on the boundary conditions and on the precise manner
in which the microwaves are coupled into the system.
Such a calculation is beyond the scope of this work.
However, if the spatial variation of the field is
sufficiently slow (i.e. $q_0$ is sufficiently small)
the field inside the metal  can still be approximated by a plane 
wave.
To estimate the upper limit for $q_0$ where this approximation
is correct, let us assume that the external potential
applied to the electrons is
$V^{\rm ext} ( {\bf{r}} , t ) = V_0^{\rm ext}  
\sin ( {\bf{q}}_0 \cdot {\bf{r}} 
- \omega_0 t )$. 
The total potential is then
 \begin{equation}
 V ( {\bf{r}} , t ) = V_0^{\rm ext} {\rm Im}
\left[ 
\epsilon^{-1} ( {\bf{q}}_0 , \omega_0 ) 
 e^{ i  ( {\bf{q}}_0 \cdot {\bf{r}} - \omega_0 t ) } 
\right]
\; ,
\end{equation}
where the longitudinal electric function
in the diffusive regime is given in Eq.(\ref{eq:epsdef}).
Screening can be ignored in the regime
where $\epsilon ( {\bf{q}}_0 , \omega_0 )$ can be approximated by
unity. From Eq.(\ref{eq:epsdef}) 
it is easy to see that in $d=1$ this is the case
when
\begin{equation}
\label{screencond1}
 q_0 \ll \sqrt{ \frac{\omega_0}{D}} \; ,
\end{equation}
while in $d=2$  
the external field is effectively not screened if
\begin{equation}
\label{screencond2}
q_0 \ll \frac{\omega_0}{2 \pi e^2 \nu_2 D} \; .
\end{equation}
Keeping in mind that in a good metal
$2\pi e^2\nu_2 > 1/ \ell$, we see that Eqs.(\ref{screencond1},\ref{screencond2})
together with $ \omega_0 < 1 / \tau_{\rm el}$
are more restrictive than the condition
$q_0<1/ \ell$   which has to be satisfied 
in order to use the semiclassical Eq.(\ref{eq:Cooperondef}).
We  assume that $q_0$ is sufficiently small
so that the inequalities (\ref{screencond1},\ref{screencond2}) are valid.
From Eq.(\ref{Vt0}) we
then obtain 
 \begin{eqnarray}
 V_{t_0} ( {\bf{q}} , \omega ) = \frac{V_0^{\rm ext}}{2}(2\pi)^{d+1} 
\left[ \delta(\omega-\omega_0) - \delta(\omega + \omega_0) \right] 
\nonumber \\
\times \left[ e^{ -i\omega_0 t_0  } \delta({\bf q - q}_0) + e^{
    i\omega_0 t_0  } \delta({\bf q + q}_0)    
\right]
\; .
\end{eqnarray}
This expression is now inserted into the general results given in
Eqs.(\ref{eq:F1sol}) and (\ref{eq:F2sol}). For $F_1$ we get
\begin{eqnarray}
\label{F1expl}
F_1({\bf k},\tau ; {\bf r},t) = - i \frac{e E_0}{ q_0} \biggl\{e^{i({\bf
    q}_0\cdot {\bf r}-\omega_0 t_0)} {\text {Im}
  }\biggl(\frac{1-e^{-D({\bf q}^2_0 +2{\bf k\cdot q}_0)\tau}
  e^{i\omega_0\tau}}{D({\bf 
    q}^2_0 +2{\bf k\cdot q}_0) -i\omega_0} e^{-i\omega_0 t} \biggr)
\nonumber \\
+ e^{-i({\bf
    q}_0\cdot {\bf r}-\omega_0 t_0)} {\text{Im}
  }\biggl(\frac{1-e^{-D({\bf q}^2_0 -2{\bf k\cdot q}_0)\tau}
  e^{i\omega_0\tau}}{D({\bf 
    q}^2_0 -2{\bf k\cdot q}_0) -i\omega_0} e^{-i\omega_0 t} \biggr)
\biggr\} \; .
\end{eqnarray}
Performing the analogous calculation for the function $F_2$ 
given in Eq.(\ref{eq:F2sol})
would result in a rather lengthy
expression. 
Since in the following we are only interested in the leading
terms of an expansion in
powers of $q_0$, it is more convenient to first expand $F_1$
and then use Eq.(\ref{F1F2}) to obtain $F_2$. 
 
If we take the limit ${\bf{q}}_0  \rightarrow 0$ keeping 
$E_0$ constant we obtain a spatially 
constant field, which has been considered
in Ref.\cite{Altshuler81}.
In this limit the microwaves do not affect
the density of the electrons.
For finite
$q_0$, however, the electric field 
induces a density modulation with amplitude
$\rho_0 = E_0 q_0 / (4 \pi )$.
We now show that 
for sufficiently small microwave power $P$ this leads to a 
dephasing rate $1/ \tau_{\rm AC}$ 
proportional to $P^{1/2}$,
as observed  in Ref.\cite{Wang87}.
Note that the microwave power
$P$ coupled into the sample can be estimated as
\cite{Altshuler98,Webb99}
\begin{equation}
P=\frac{(E_0 L)^2}{2 R_{\rm tot}} \propto E_0^2 \; ,
\end{equation}
where $L$ is the effective sample length including leads, and $R_{\rm tot}$
is the total resistance.

According to Eqs.(\ref{eq:sigmane},\ref{eq:Gsol})
the weak localization correction $\delta \sigma$ 
is determined by $F ( {\bf{k}} , t ; {\bf{r}} , t/2 )$.
In the experiments\cite{Wang87,Webb99}
the wavelength of the 
microwaves is larger than the size of the sample,
so that we may expand
$F ( {\bf{k}} , t ; {\bf{r}} , t/2 )$
in powers of $q_0$.
From Eq.(\ref{F1expl}) we obtain
 \begin{eqnarray}
 F_1 ( {\bf{k}} , t ; {\bf{r}} , t/2 )
 & = &     2 D e E_0 \omega_0^{-2 }
 g_1 ( \omega_0 t ) \{
 2 k_{\|} 
  \sin ( \omega_0 t_0 )  
 \nonumber
 \\
 &  & \hspace{-24mm}  + 
 q_0 
 \cos ( \omega_0 t_0 ) 
 [ {i} 
 ( 1  -  2 D k_{\|}^2 t
 )
 +   2 {{r}}_{\|} 
  k_{\|} 
 ]
 \}
  +   O ( E_0 q_0^2 )
 \; .
 \label{eq:F1q}
\end{eqnarray}
where $k_{\|} = {\bf{k}} \cdot {\bf{q}}_0 / q_0$,
$r_{\|} = {\bf{r}} \cdot {\bf{q}}_0  / q_0$ and
\begin{equation}
 g_1 ( x )  =  x  \cos \left(\frac{x}{2}\right) -   2 \sin \left(
   \frac{x}{2} \right)  \; .
\end{equation}
Inserting this result into Eq.(\ref{F1F2}) yields
\begin{eqnarray}
 F_2 ( {\bf{k}} , t ; {\bf{r}} , t/2 ) & = &
  4 D ( e E_0 )^2 \omega_0^{-3} 
 g_{2} ( \omega_0 t )
 \sin^2 ( \omega_0 t_0 )
 \nonumber 
 \\
 & + & 
 O ( E_0^2 q_0 )
 \; ,
 \label{eq:F2q}
 \end{eqnarray}
with
\begin{equation}
 g_2 ( x ) = x   + \frac{x}{2} \cos (x) 
 - \frac{3}{2} \sin (x) \; .
\end{equation}
Note that the second line in Eq.(\ref{eq:F1q}) is 
the leading correction for finite $q_0$ and small $E_0$.
It can be shown\cite{Voelker99}
that in the limit $q_0 \rightarrow 0$ all
higher order terms $F_n$, $ n \geq 3$ in our
eikonal expansion vanish, so that the
exact solution of the eikonal equation
(\ref{eq:F}) is given by
$F_{q_0 = 0} = [F_1 + F_2 ]_{q_0 = 0}$.
The fact that the Cooperon in
a spatially constant field can be calculated
exactly is obvious in a gauge
where the electric field is represented
in terms of a vector potential\cite{Altshuler81}.
It is reassuring to see that
our eikonal expansion reproduces
the exact solution in a different gauge.
Substituting Eqs.(\ref{eq:F1q},\ref{eq:F2q}) into
Eq.(\ref{eq:Gsol}) and performing the  ${\bf{k}}$-integration, 
we obtain for the weak localization correction to
the static conductivity
 \begin{eqnarray}
 \delta \sigma
  &  = & - \frac{ \sigma_0 }{\pi \nu_d } 
  \frac{ \omega_0^{ \frac{d}{2} -1 } }{ ( 4 \pi D)^{d/2}} 
 \int_{ \omega_0 \tau_{\rm el} }^{\infty} \frac{d x }{x^{d/2}}
 e^{ - \alpha g_3 ( x  )  - \gamma x}
 \nonumber
 \\
 & \times & \frac{1}{\pi}
 \int_{0}^{   \pi} d \varphi 
 \frac{ e^{ \alpha g_3 ( x ) \cos  ( 2  \varphi )  
 - i \beta g_1 ( x ) \cos  \varphi }}{
 \sqrt{ 1 - 2 i \beta g_1 ( x ) \cos  \varphi   } }
 \; .
 \label{eq:Gdabc}
 \end{eqnarray}
where
\begin{equation}
g_3 (x) = x + \sin x - \frac{8}{x} \sin^2 \left(\frac{x}{2}\right) \;
,
\label{eq:g3def}
\end{equation} 
and we have introduced the dimensionless parameters 
 \begin{eqnarray}
 \alpha & = & \frac{D ( e E_0 )^2 }{ \omega_0^3}
 \; ,
 \label{eq:a1}
 \\
 \beta & = & \frac{ 2 D | e E_0 | q_0  }{ \omega_0^2}
 \; ,
 \label{eq:b1}
 \\
\gamma & = &  \frac{ \Gamma_0 }{ \omega_0}
 \; .
 \label{eq:c1}
 \end{eqnarray}
Note that after the $\varphi$-integration 
the integral in Eq.(\ref{eq:Gdabc}) is real, so that
we may replace the integrand by its real part, which 
is independent of the sign of $E_0$. 
This is the reason why in the definition (\ref{eq:b1}) of $\beta$
only the absolute value of  $E_0$ appears. 
Note that $\beta$ is proportional to the absolute value of the
amplitude $\rho_0 = E_0 q_0 / (4 \pi)$ of the
density wave associated with the longitudinal
electric field.

The order of magnitude of the non-equilibrium dephasing rate 
$1 / \tau_{\rm AC}$ can be estimated  as
$1/\tau_{\rm AC} \approx \omega_0 /x_c$, where
$x_c$ is the effective
large-$x$ cutoff 
for the integration in Eq.(\ref{eq:Gdabc}).
For simplicity  we now set
$\gamma = 0$.
Depending on the ratio
$\alpha / \beta = | e E_0 | /  ( 2 \omega_0 q_0 ) 
= | V_0 | /  ( 2 \omega_0)$, we obtain different
behaviors for $1/ \tau_{\rm AC}$.
In the limit $\alpha / \beta \rightarrow \infty$ 
we recover the results of
Ref.\cite{Altshuler81}.  In this case
$1/ \tau_{\rm AC} \propto \omega_0 \alpha$
for $\alpha \ll 1$, and
$1/ \tau_{\rm AC} \propto \omega_0 \alpha^{1/5}$
for $\alpha \gg 1$.
Note that $\alpha$ is proportional the
microwave power $P \propto E_0^2$ absorbed by the system,
while $\beta$ is proportional to $P^{1/2}$.
Obviously, for sufficiently small
$P$ we always have  $ \beta \gg \alpha$.
Then the cutoff 
for the $x$-integration in Eq.(\ref{eq:Gdabc})
is determined by $\beta$, and we can set $\alpha=0$ to determine
$x_c$. The $x$-integration is effectively cut of where the integrand
starts to oscillate. Thus we estimate $x_c$ from the condition $\beta
g_1(x_c) \approx 1$. For $\beta\ll 1$ the effective cutoff 
$x_c$ is large compared with unity, so that with Eq.(\ref{eq:g3def})
we obtain $1/x_c \propto \beta$. 
This implies
\begin{equation}
\frac{1}{\tau_{\rm AC }} \propto \omega_0\beta \; \; , \; \;  \text{for}\;
\alpha\ll\beta\ll 1 \; .
\end{equation}
On the other hand, for $\beta\gg 1$ the $x$-integration is cut off at
small $x$ and we approximate $g_1(x)\approx -\frac{x^3}{12}$. This
yields
\begin{equation}
\frac{1}{\tau_{\rm AC}} \propto \omega_0\beta^{1/3} \; \; , \; \;  \text{for}\;
\alpha\ll\beta \; \text{and} \; \beta\gg 1 \; .
\end{equation}
Hence, for sufficiently small $P$ the dephasing rate
$1/ \tau_{\rm AC}$ is proportional to
$P^{1/2}$. Moreover, in this regime
$\omega_0 / \tau_{\rm AC}$ should
be independent of $\omega_0$ 
as long as
the dispersion of the longitudinal density wave
(i.e. the dependence of $q_0$ on $\omega_0$)
can be neglected.
Note that the condition $\beta \gg \alpha \gg \gamma$
where $1/ \tau_{\rm AC}$ should exhibit a $P^{1/2}$-dependence
can also be written as
 \begin{equation}
 D q_0^2 \gg \frac{\omega_0}{ | V_0 | } \Gamma_0
 \gg \Gamma_0
 \; .
 \label{eq:condition}
 \end{equation}
Thus, for finite $\Gamma_0$ the value of $q_0$ must be sufficiently large
to observe the $P^{1/2}$-law. 
Keeping in mind that according to the conventional point of 
view\cite{Aleiner98,Altshuler98,Altshuler81} the intrinsic
dephasing rate $\Gamma_0$ should vanish for $T \rightarrow 0$,
we conclude that
at sufficiently low temperatures and small $V_0$
Eq.(\ref{eq:condition})  can be satisfied for
experimentally relevant wave-vectors $q_0$.
Note that according to Ref.\cite{Altshuler98}
the observed saturation of the dephasing time, which is typically
of the order of a few nanoseconds, is due to 
some external noise. If this is correct, then 
the experimentally observed saturation value of the dephasing rate
should not be identified with $\Gamma_0$.
The same arguments apply for the condition $| V_0 | \gg\Gamma_0$ that
follows from Eq.(\ref{eq:condition}) together with
Eqs.(\ref{screencond1},\ref{screencond2}). For low enough temperatures
this inequality should always be satisfied. (Most recent experimental
data \cite{Gougam99} indicates $\Gamma_0\rightarrow 0$ for $T\rightarrow
0$ in narrow Ag wires. These probes therefore seem to be suiting to
test our findings.) In conclusion, in the
limit $T\rightarrow 0$, the only relevant restriction following from
Eq.(\ref{eq:condition}) should be $ | V_0 | \ll   \omega_0$ .

For convenience, 
let us summarize the assumptions made in our derivation of the
dephasing time due to external
microwave radiation in this section.
First of all, we have assumed that the microwave field
inside the metal can be approximated by
a propagating wave with wave-vector ${\bf{q}}_0$,
frequency $\omega_0$ and amplitude
$E_0 = q_0 V_0 / e$. 
We have argued that this approximation
is valid in a regime where the dielectric function
of the bulk system can be approximated by unity,
which is possible if
$q_0\ll \sqrt{\omega_0/D}$ in $d=1$, and
$q_0\ll \omega_0/(2\pi e^2\nu_2 D)$ in $d=2$.
Throughout this work we have assumed 
diffusive dynamics and calculated the 
weak localization correction to the conductivity
from the  semiclassical equation (\ref{eq:Cooperondef}),
which is of course only
correct for sufficiently small wave-vectors ($q_0 < 1 / \ell$)  and
frequencies ($\omega_0 < 1 / \tau_{\rm el}$).
Note that the latter inequality sets the upper
limit for the microwave frequency 
where our approach remains valid.
If these restrictions are satisfied, then
the dephasing rate due to microwaves  is
\begin{equation}
\frac{1}{\tau_{\rm AC}} \propto 
\left\{
\begin{array}{ll}
\omega_0^{2/5} P^{1/5} & \text{for} \; | V_0 | \gg  \omega_0 \; \text{and} \;
D ( e E_0 )^2 \gg \omega_0^2 \; ,
\\
\omega_0^{-2} P & \text{for} \;   | V_0 | \gg \omega_0
\; \text{and} \; D ( e E_0 )^2 \ll \omega_0^2 \; ,
\\
(q_0\omega_0)^{1/3} P^{1/6} & \text{for} \;  | V_0 | \ll \omega_0
\; \text{and} \; D | e E_0 | q_0 \gg \omega_0^2  \; ,
\\
q_0 \omega_0^{-1} \sqrt{P} & \text{for} \;  | V_0 | \ll \omega_0
\; \text{and} \; D | e E_0 | q_0 \ll \omega_0^2  \; ,
\end{array}
\right.
\end{equation}
where  $P \propto E_0^2$ is the
microwave-power absorbed by the system, and
$V_0 = e E_0 / q_0$.

\section{Comparison with experiments and summary}

\begin{figure}
\epsfysize8cm 
\hspace{5mm}
\epsfbox{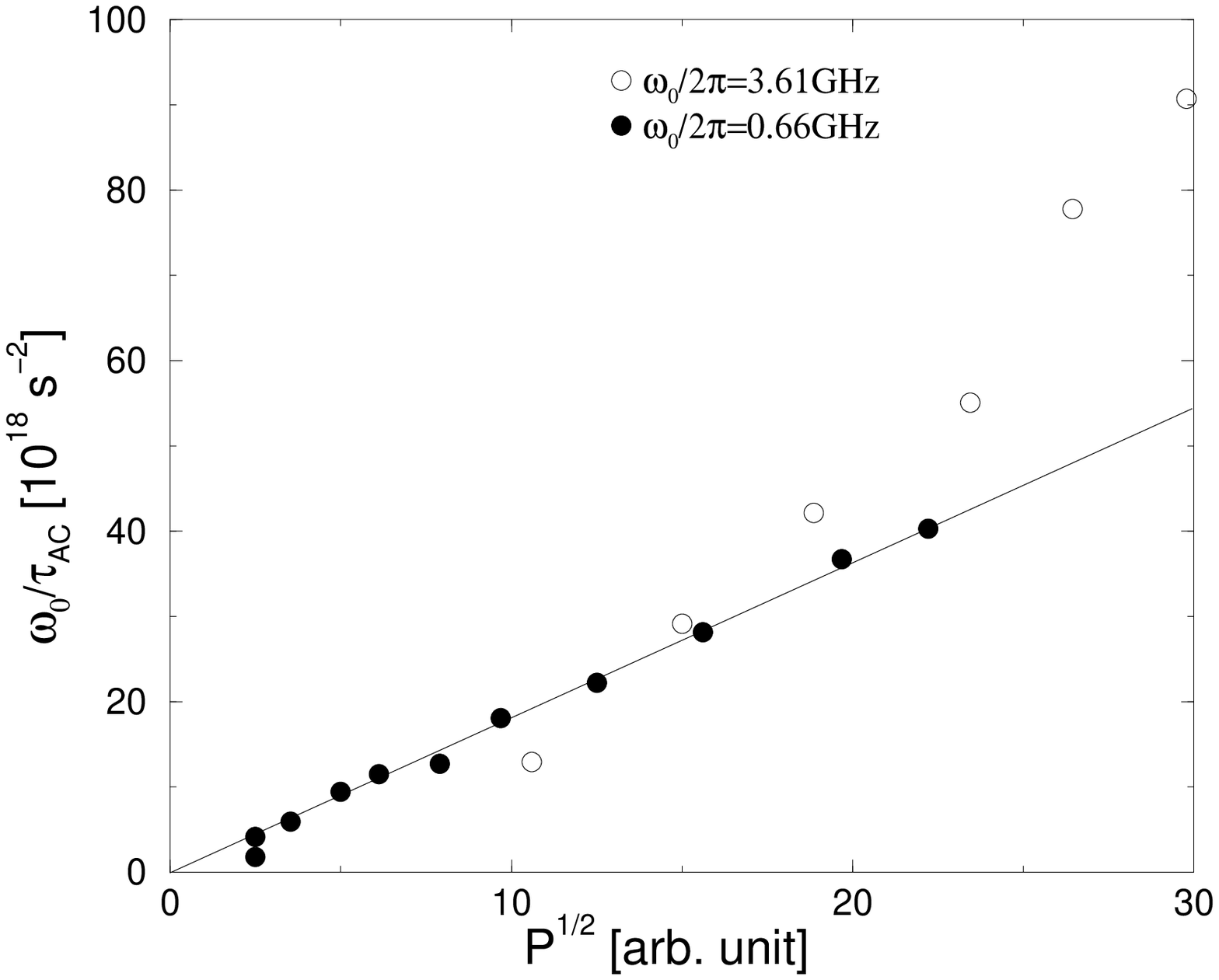}
\vspace{5mm}
\caption{
Data for $ \omega_0 / \tau_{\rm AC}$
as function of $P^{1/2}$
from Fig.3 of Ref.[7(a)].
Here $P$ is the 
microwave power  coupled into the system. 
The solid line is a fit of the data 
at $\omega_0 / 2 \pi = 0.66 {\rm GHz}$
to our prediction
$\omega_0 / \tau_{\rm AC}
\propto P^{1/2}$, which is valid for small $P$.
}
\label{fig:Wang}
\end{figure}

Some time ago Wang and Lindelof\cite{Wang87}
have measured $1/\tau_{\rm AC}$ as a function of $P$
in magnesium films.
Their data from Ref.[7(a)] are reproduced in 
Fig.\ref{fig:Wang}.
At $\omega_0 / 2 \pi = 0.66 {\rm GHz}$
our prediction $\omega_0 / \tau_{\rm AC}
\propto P^{1/2}$ is in good agreement  with the experiment. 
Although the data
at $\omega_0 / 2 \pi = 3.61 {\rm GHz}$
cannot be fitted by a straight line through the origin,
the data roughly scale as
$1/ \tau_{\rm AC} \propto \omega_0^{-1}$
for fixed and small $P$.
One should keep in mind, however, that in the experiments\cite{Wang87}
the precise value of the microwave power coupled into the
system was not measured, and the power axis
for the two sets of data was rescaled differently.
Further evidence for 
the $P^{1/2}$-law can be found in Fig.19 of 
Ref.[7(b)].

Recent measurements of $1/ \tau_{\rm AC}$  
by Webb {\it{et al.}}\cite{Webb99}
suggest  a $P^{1/5}$-law  
for large $P$ in
a limited range of frequencies. 
Our calculation shows that the
$P^{1/5}$-law should hold as long as 
$\alpha  \gg {\rm max} \{ \beta , 1 \}$, while
for $\beta  \gg {\rm max} \{ \alpha , 1 \}$
we predict
$1/ \tau_{\rm AC} \propto \beta^{1/3} \propto P^{1/6}$.
Keeping in mind that $\alpha / \beta \propto \omega_0^{-1}$
we predict a crossover from a $P^{1/5}$- via a $P^{1/6}$-
to a $P^{1/2}$-behavior as the frequency is increased.
The data shown in Fig.6 of Ref.\cite{Webb99}
are consistent with the existence of such a crossover.

In summary, by means of an eikonal expansion for the Cooperon
in a slowly varying  scalar potential we have
derived a general
expression for the dephasing rate 
due to fluctuating electric fields in disordered
metals. Our method is a physically transparent
alternative to the path-integral
approach used in Ref.\cite{Altshuler82}. 
For randomly fluctuating fields
with zero average and arbitrary covariance
the dephasing rate can be obtained from
Eqs.(\ref{eq:Qares},\ref{eq:Qbres}). 
These expressions 
take vertex corrections into account
and remain finite in reduced dimensions,
where the well-known\cite{Chakravarty86,Imry99}
semiclassical result Eq.(\ref{eq:Qb2})
is infrared divergent for some physically relevant 
$g ( {\bf{q}} , \omega)$.
For example, in the case of $1/f$-noise\cite{Imry99},
where $g ( {\bf{q}} , \omega ) \propto 1 / \omega$,
the integral in Eq.(\ref{eq:Qb2}) 
is divergent in $d \leq 2$,
while  our more general result 
(\ref{eq:Qares},\ref{eq:Qbres}) is finite.
We have also studied dephasing due to external electric fields, 
and have proposed 
an explanation for the data of Ref.\cite{Wang87}.
Finally, we point out that
our eikonal method 
can  also be used to calculate
the Diffuson in a fluctuating electric
field, which is of current interest
in several contexts\cite{Polyakov98}.

\section* {Acknowledgments}

We thank P. E. Lindelof and P. Mohanty for their comments
and for helping
us to understand the experiments\cite{Wang87,Webb99}.
This work was supported by the
DFG via the  Heisenberg Programm and 
SFB 345.

%

%
%

\end{document}